\documentclass[%
 aip,
 amsmath,amssymb,
 reprint,%
]{revtex4-1}

\usepackage{graphicx}
\usepackage{dcolumn}
\usepackage{bm}

\usepackage[utf8]{inputenc}
\usepackage[T2A]{fontenc}
\usepackage{etoolbox}
\usepackage[main=english,russian]{babel}
\usepackage{xcolor}

\draft 

\begin{document}

\title{Magnetic stagnation of two counterstreaming plasma jets induced by intense laser } 

\author{R. S. Zemskov}
\email[]{roman.zemskov.6@gmail.com}
\affiliation{A.V. Gaponov-Grekhov Institute of Applied Physics of the Russian Academy of Sciences (IAP RAS), 46 Ulyanov st., Nizhny Novgorod 603950, Russia}

\author{S. E. Perevalov}
\affiliation{A.V. Gaponov-Grekhov Institute of Applied Physics of the Russian Academy of Sciences (IAP RAS), 46 Ulyanov st., Nizhny Novgorod 603950, Russia}

\author{A. V. Kotov}
\affiliation{A.V. Gaponov-Grekhov Institute of Applied Physics of the Russian Academy of Sciences (IAP RAS), 46 Ulyanov st., Nizhny Novgorod 603950, Russia}
 
  \author{A.A. Murzanev}
\affiliation{A.V. Gaponov-Grekhov Institute of Applied Physics of the Russian Academy of Sciences (IAP RAS), 46 Ulyanov st., Nizhny Novgorod 603950, Russia}

\author{A. I. Korytin}
\affiliation{A.V. Gaponov-Grekhov Institute of Applied Physics of the Russian Academy of Sciences (IAP RAS), 46 Ulyanov st., Nizhny Novgorod 603950, Russia}

 \author{K. F. Burdonov}
\affiliation{A.V. Gaponov-Grekhov Institute of Applied Physics of the Russian Academy of Sciences (IAP RAS), 46 Ulyanov st., Nizhny Novgorod 603950, Russia}

 \author{V.N. Ginzburg}
\affiliation{A.V. Gaponov-Grekhov Institute of Applied Physics of the Russian Academy of Sciences (IAP RAS), 46 Ulyanov st., Nizhny Novgorod 603950, Russia}

 \author{A.A. Kochetkov}
\affiliation{A.V. Gaponov-Grekhov Institute of Applied Physics of the Russian Academy of Sciences (IAP RAS), 46 Ulyanov st., Nizhny Novgorod 603950, Russia}

 \author{S.E. Stukachev}
\affiliation{A.V. Gaponov-Grekhov Institute of Applied Physics of the Russian Academy of Sciences (IAP RAS), 46 Ulyanov st., Nizhny Novgorod 603950, Russia}

 \author{I.V. Yakovlev}
\affiliation{A.V. Gaponov-Grekhov Institute of Applied Physics of the Russian Academy of Sciences (IAP RAS), 46 Ulyanov st., Nizhny Novgorod 603950, Russia}

 \author{I.A. Shaikin}
\affiliation{A.V. Gaponov-Grekhov Institute of Applied Physics of the Russian Academy of Sciences (IAP RAS), 46 Ulyanov st., Nizhny Novgorod 603950, Russia}

 \author{A.A. Kuzmin}
\affiliation{A.V. Gaponov-Grekhov Institute of Applied Physics of the Russian Academy of Sciences (IAP RAS), 46 Ulyanov st., Nizhny Novgorod 603950, Russia}

 \author{E.V. Derishev}
\affiliation{A.V. Gaponov-Grekhov Institute of Applied Physics of the Russian Academy of Sciences (IAP RAS), 46 Ulyanov st., Nizhny Novgorod 603950, Russia}

 \author{A.V. Korzhimanov}
\affiliation{A.V. Gaponov-Grekhov Institute of Applied Physics of the Russian Academy of Sciences (IAP RAS), 46 Ulyanov st., Nizhny Novgorod 603950, Russia}

\author{A. A. Soloviev}
\affiliation{A.V. Gaponov-Grekhov Institute of Applied Physics of the Russian Academy of Sciences (IAP RAS), 46 Ulyanov st., Nizhny Novgorod 603950, Russia}

 \author{A.A. Shaykin}
\affiliation{A.V. Gaponov-Grekhov Institute of Applied Physics of the Russian Academy of Sciences (IAP RAS), 46 Ulyanov st., Nizhny Novgorod 603950, Russia}

 \author{A. N. Stepanov}
\affiliation{A.V. Gaponov-Grekhov Institute of Applied Physics of the Russian Academy of Sciences (IAP RAS), 46 Ulyanov st., Nizhny Novgorod 603950, Russia}


\author{M. V. Starodubtsev}
\affiliation{A.V. Gaponov-Grekhov Institute of Applied Physics of the Russian Academy of Sciences (IAP RAS), 46 Ulyanov st., Nizhny Novgorod 603950, Russia}

\author{E. A. Khazanov}
\affiliation{A.V. Gaponov-Grekhov Institute of Applied Physics of the Russian Academy of Sciences (IAP RAS), 46 Ulyanov st., Nizhny Novgorod 603950, Russia}

\date{\today}

\begin{abstract}
Experiments with interacting high-velocity flows of laser plasma can help answer the fundamental questions in plasma physics and improve the understanding of the mechanisms behind astrophysical phenomena, such as formation of collisionless shock waves, deceleration of accretion flows, and evolution of solar (stellar) flares. This work presents the first direct experimental observations of stagnation and redirection of counterstreaming flows (jets) of laser plasma induced by ultra-intense laser pulses with intensity $I \sim$ 2 $\times$ $10^{18}$ $W/cm^2$. 
Hybrid (PIC - fluid) modeling, which takes into account the kinetic effects of ion motion and the evolution of the pressure tensor for electrons, demonstrates the compression of counterdirected toroidal self-generated magnetic fields embedded in the counterstreaming plasma flows. The enhancement of the toroidal magnetic field in the interaction region results in plasma flow stagnation and redirection of the jets across the line of their initial propagation.


\end{abstract}

\maketitle 

\section{\label{sec1}Introduction}

Counterstreaming plasma flows are observed in astrophysical objects and, in some cases, lead to the formation of highly inhomogeneous nonequilibrium plasma structures. The well-known examples include solar wind interaction with planetary magnetospheres, colliding stellar winds \cite{eichler1993particle, reimer2006nonthermal, mikic2006introduction, chen2017physics}, supernova remnants (see, e.g., Ref. \cite{Vink2012AA} for a review), pulsar wind nebulae (see, e.g., Ref.\cite{kargaltsev2015pulsar} for a review), gamma-ray burst afterglows \cite{miceli2022gamma}, radilobes of active galactic nuclei\cite{wilson2001chandrax, croston2007shock}, deceleration of accretion flows  \cite{gilfanov2014radiation, vincentelli2023shared}. While the global hydrodynamics of these shocks is relatively well understood, the details of their microphysics are still an open issue \cite{marcowith2016microphysics}. The notable unsolved problems are properties of small-scale magnetic fields (magnetic turbulence) and the process of initial particle injection into non-thermal distribution, which further undergoes diffusive shock acceleration. Recent X-ray polarization measurements \cite{ferrazzoli2023x, prokhorov2024evidence} provide a glimpse on the first problem, whereas the second can only be studied by means of either numerical models or laboratory modeling. Microphysics of collisionless shocks is a complex problem involving plenty of plasma instabilities and various elementary processes. It is not trivial even to list all the relevant physical phenomena, much less to implement them in numerical models. Laboratory experiments with plasma flows complement numerical simulations in revealing the details of collisionless-shock microphysics.



Besides, the role of the large-scale magnetic field in the microphysics of collisionless shock waves is still not fully understood. The research and interpretation of results are complicated by the fact that, in addition to the external magnetic field, the dynamics of plasma flows can also be influenced by the self-induced large-scale magnetic field. It is known that without any seed a magnetic field can be generated through several mechanisms, such as the Biermann battery \cite{doi2011generation} or as a result of the development of Weibel-type instabilities in the plasma \cite{weibel1959spontaneously}. The counterstreaming plasma flows, in which a self-generated toroidal magnetic field is embedded, can lead to entirely new effects, such as redirection of plasma flows, magnetic reconnection \cite{fiksel2014magnetic, hesse2020magnetic}, and plasma stagnation \cite{li2013structure, ryutov2013magnetic}.

A large distance to the cosmic objects with counterstreaming jets under consideration complicates direct observation of flow redirection and testing of theoretical models. However, laboratory approaches offer a possibility of creating controlled experimental conditions, allowing detailed investigation of the evolution of plasma structures. Experiments enable the control of key parameters that are difficult or impossible to measure in observations of cosmic objects, which in turn contributes to a deeper understanding of the physical mechanisms behind the observed phenomena. Thus, laboratory experiments are a crucial companion to astronomical observations \cite{remington1999modeling}.

Modern laser systems allow creating plasma flows with fairly high directed velocities (on the order of and exceeding 100 km/s), which enables collisionless ions in counterpropagating flows, while electrons and ions in each flow remain collisional \cite{ryutov2014collisional}. Therefore, a magnetohydrodynamic approach can be used for an approximate description of the evolution of individual plasma flows \cite{ryutov2013magnetic}, but for an accurate description of the interaction between the two flows, the kinetics of the ions must be taken into account.

Experiments aimed at investigating collision of quasi counterstreaming plasma flows have primarily been conducted using energetic (several kJ) and moderately intense $I < 10^{15}$ W/cm$^2$ nanosecond laser pulses \cite{park2012studying,li2013structure, kugland2013visualizing, fiksel2014magnetic, morita2010collisionless, morita2020local, morita2019anomalous, woolsey2004laboratory, yuan2017formation, fazzini2022particle, morita2013interaction, kuramitsu2011time, courtois2004experiment, ross2012characterizing}, as well as on wire-array facilities \cite{morita2020local, russell2022perpendicular}. Experiments like ours where counterstreaming plasma flows are created by irradiating solid targets with ultra-intense short (a few tens of fs to a few ps) laser pulses with peak intensities exceeding 10$^{18}$ W/cm$^2$ have been absent until now. This significantly new approach was proposed in the theoretical work \cite{fiuza2012weibel} for generating collisionless shock waves. When plasma is heated with short ultra-intense pulses, there are several significant differences compared to nanosecond plasma heating. Under femtosecond intense irradiation of the target, a substantial fraction of accelerated electrons is generated, with oscillatory energies up to few MeV \cite{antici2013modeling}. Energetic electrons can escape and move away from the heating region \cite{dubois2014target}, resulting in an huge electric field that returns electrons and accelerates ions in some cases to 10 and even 100 MeV \cite{soloviev2017experimental, daido2012review}. In this process, the direct and reverse oscillating electron flows contribute to the generation of magnetic fields, such as large-scale fountain magnetic fields \cite{sarri2012dynamics, albertazzi2015dynamics, shaikh2016megagauss, borghesi1998large, gopal2008temporally} and small-scale filamentation fields arising from the development of Weibel-type current instabilities \cite{weibel1959spontaneously, ruyer2020growth, krishnamurthy2020observation, quinn2012weibel}. Fountain toroidal self-generated magnetic fields in plasma can exceed Biermann fields and reach several hundred tesla \cite{shaikh2016megagauss, borghesi1998large, gopal2008temporally}, and even more than 30,000 T reported in some cases \cite{tatarakis2002measurements, tatarakis2002measuring}. These magnetic fields can exert a constraining influence on the radial expansion of the plasma \cite{sarri2012dynamics}, leading to a collimated plasma flow, which we call a jet.

Our work is dedicated to the experimental investigation of the interaction of counterstreaming plasma jets generated by ultra-intense laser pulses (I $\sim$ 2 $\times$ 10$^{18}$ W/cm$^2$). We developed and implemented an original scheme that allowed us to split the impact of the ultra-intense laser pulse across several targets and generate counterstreaming plasma jets. All experiments were conducted on the single-channel petawatt PEARL laser facility  \cite{lozhkarev2007compact, ginzburg202111, soloviev2024research}. A detailed study of the topology and evolution of the interacting plasma flows and magnetic fields was carried out using two-dimensional interferometric plasma density measurements, shadowgraphy and Faraday polarimetry measurements with femtosecond temporal resolution. Numerical computations performed using a hybrid (PIC - fluid) approach confirmed the mechanism of plasma stagnation and redirection by the toroidal self-generated magnetic field, which is advected and amplified in the interaction region of the flows.

This work does not aim to elucidate the mechanism of generating a toroidal large-scale magnetic field, which may be related to the Biermann mechanism or the fountain effect. Instead, we focus on describing the experimentally observed phenomenon of stagnation of counterstreaming plasma jets and their radial redirection, for which the advection of the toroidal magnetic field plays a significant role. To the best of our knowledge, such studies have not been conducted so far, primarily due to the lack of multichannel petawatt laser systems.

The paper is organized as follows. Section \ref{sec:Methods} describes the setup and experimental scheme. Section \ref{sec:Res} presents the results of the experiment. Section \ref{sec:Sym} provides the results of numerical modeling using the hybrid code AKA52. Section \ref{sec:Discus} presents a discussion of the obtained results and the interpretation of the numerical computations. In Sections \ref{sec:Discus} and \ref{sec:Conclude}, we discuss the results and draw conclusions.

\section{\label{sec:Methods}Methods}

The experiments were conducted at the petawatt PEARL laser facility \cite{lozhkarev2007compact, ginzburg202111}. Details about the capabilities of the laser facility can be found in the review \cite{soloviev2024research} and in other works \cite{soloviev2022improving, burdonov2021inferring, zemskov2024laboratory, burdonov2022laboratory}. In the experiment, a laser pulse with an energy of 10-15 J, a duration of 60-80 fs, and a wavelength of 910 nm was used to generate plasma. The laser pulse was focused using an F/40 spherical mirror into a spot with a diameter approximately 100 $\mu$m. As a result, an intensity of about 2 $\times$ 10$^{18}$ W/cm$^{2}$ was achieved on the target. The intensity of the prepulse did not exceed  10$^{-7}$ relative to the main pulse.

Optical methods were used for plasma diagnostics. The diagnostic pulse was a weakened replica of the main laser pulse, with an energy of about 1 $\mu$J, a duration of 60-80 fs, and an aperture of more than 3 cm, allowing  nearly instantaneous snapshots of the entire expanding plasma. A Mach-Zehnder interferometer was used to measure plasma density (the scheme is shown in Fig. 1 in the Supplemental materials). Faraday polarimetry  \cite{swadling2014diagnosing} was employed to diagnose the generated magnetic field in the plasma (Fig. 1 in the Supplemental materials). A detailed description and approaches of the diagnostic methods are provided in the review \cite{soloviev2024research} and are included in the Supplemental materials. The snapshots of the plasma were taken from 0.5 to 50 ns after the target  irradiation. Plasma probing was conducted simultaneously in two directions using two probe laser beams, one of which is marked in the scheme in Fig. \ref{fig:Scheme}.

\subsection{\label{subsec:setup} Experimental setup }

\begin{figure}
 \includegraphics[width=8 cm]{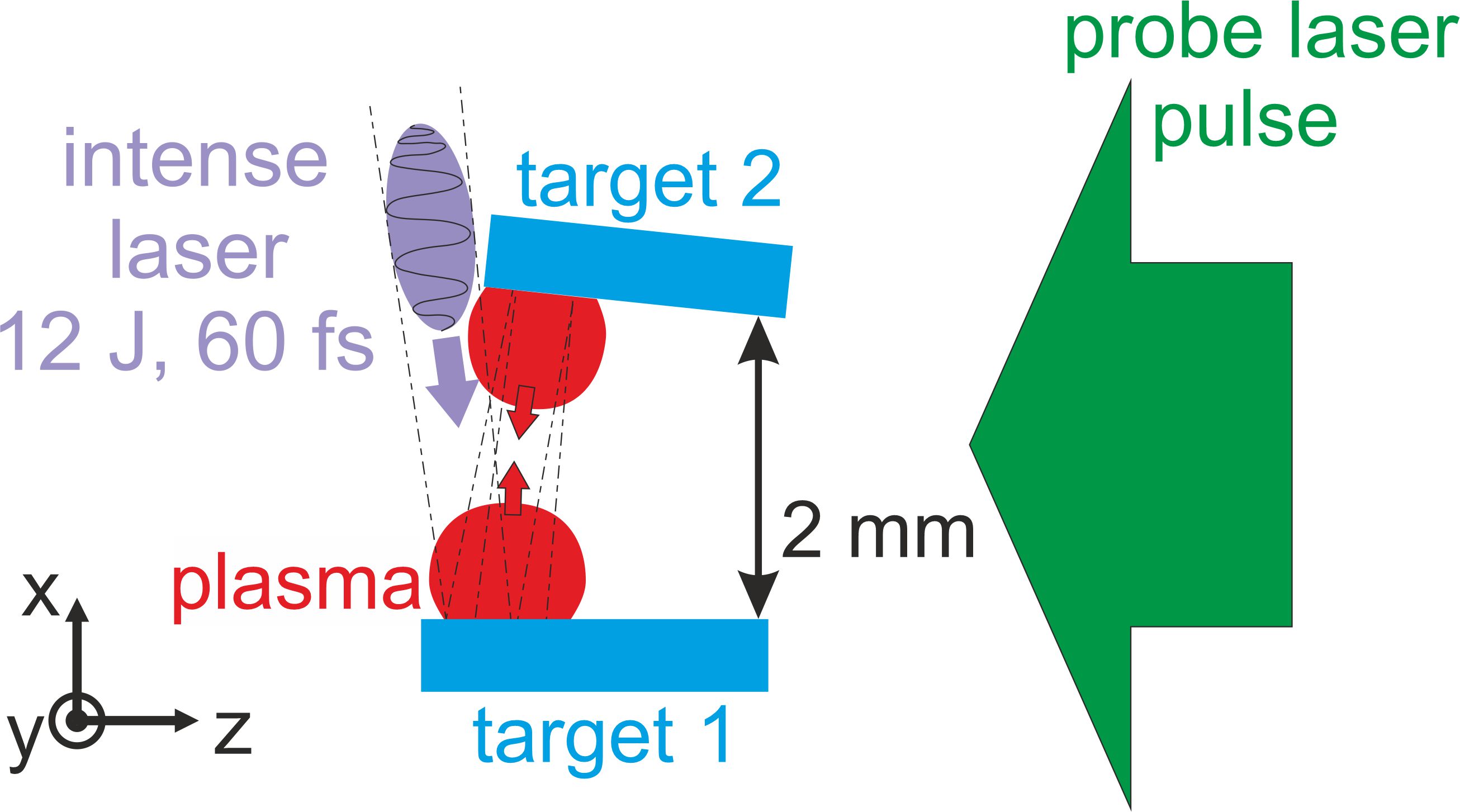}%
 \caption{\label{fig:Scheme} Experimental setup.}
\end{figure}

The experimental setup for generating counterstreaming plasma flows is shown in Fig. \ref{fig:Scheme}. A laser pulse with s-polarization ($\textbf{E}$ was in the plane of the target surface) was incident at a small angle $\theta$ of about 5$^\circ$ to the target normal, interacted with the target surface pre-ionized by the prepulse, and was reflected towards the second target located at a distance of about 2 mm. Both targets were made of the same material - polymethyl methacrylate (C$_5$O$_2$H$_8$)$_n$. The interaction of the laser pulse with the targets resulted in the generation of the laser plasma flows propagated in the directions of the target normals. The angle between the target normals was taken to be about $\theta$ = 5$^\circ$, thereby ensuring a quasi-counter collision of the two flows.

It is experimentally observed that the second plasma flow has characteristics similar to the first flow, such as density, expansion velocity, temperature, and total number of particles. This occurs due to the low absorption of intense and short laser pulses when they irradiate a solid target. Direct measurements show that a significant portion, about 70\%, of the energy of the ultra-high-power pulse is reflected from the solid target under our parameters. Thus, in our experiment, the laser pulse arriving at the second target has an energy of approximately 70\% * 12 J = 8.2 J. The Rayleigh length is about 10 mm, which significantly exceeds the 2-mm distance between the targets. Therefore, the size of the laser spot on the second target is of the same order of magnitude as the size of the spot on the first target.

\subsection{\label{subsec:OneFlow} Description of a single laser plasma flow }

\begin{figure}
 \includegraphics[width=9 cm]{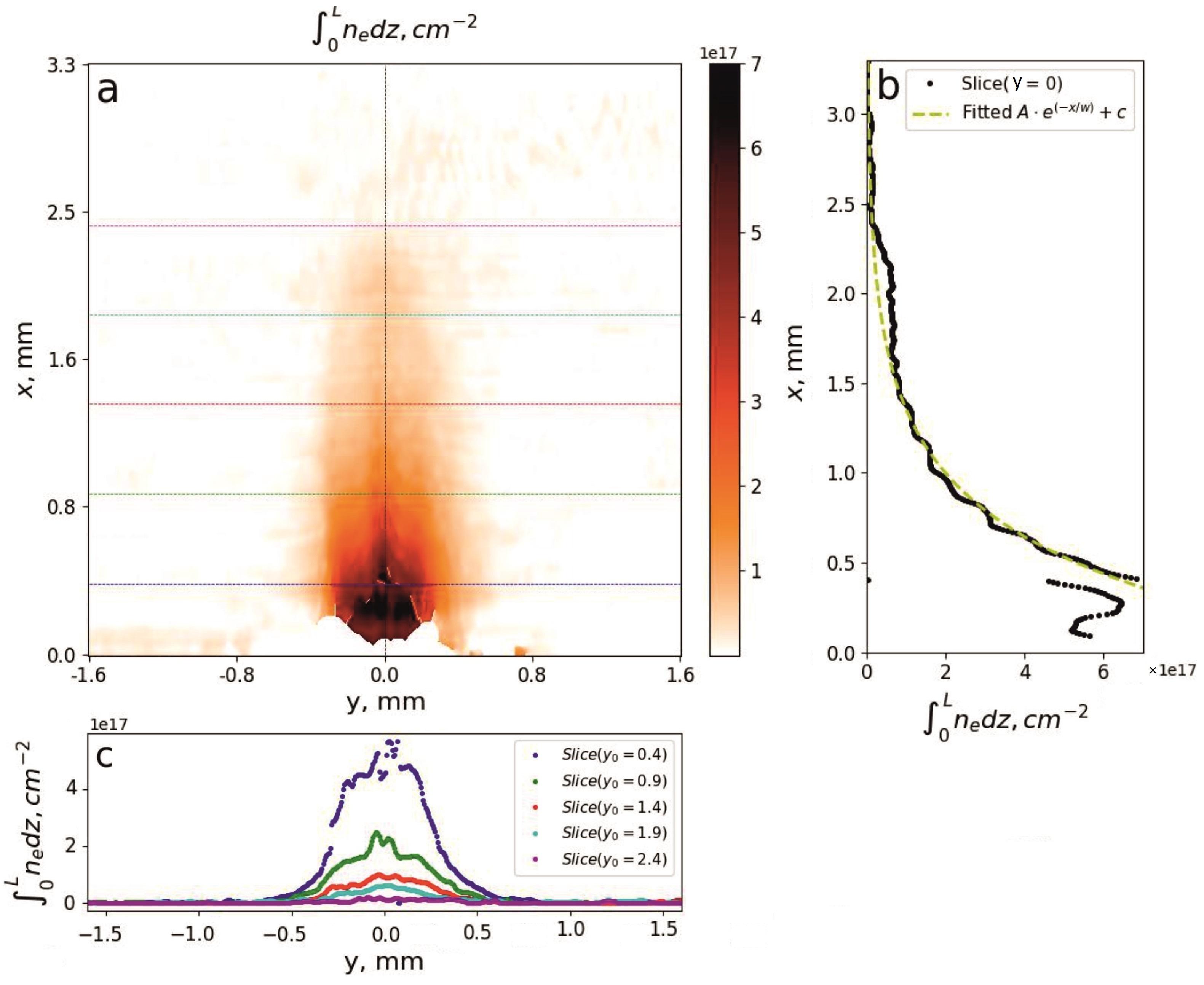}%
 \caption{\label{oneflow} Snapshot of plasma flow, obtained 8 ns after irradiating a single target with pulse intensity 2$\times 10^{18}$ W*cm$^{-2}$ (no second target and second flow). (a) Two-dimensional distribution of integrated plasma density along the beam, obtained after analyzing the interferogram. (b) Plasma density slice along x axis directed normally to target. (c) Plasma density slices along target surface.}
\end{figure}

To better understand the interaction of two plasma flows, let us first consider the characteristics of a single flow when  a single target is irradiated. In the first picoseconds after irradiation, the plasma electrons have  a very high temperature, reaching several hundred keV (see Refs. \cite{antici2013modeling, dubois2014target, soloviev2017experimental}). As the plasma is expanding, it rapidly cools down and isotropizes, resulting in a characteristic temperature of T$_e \sim$ T$_i \sim$ 50 eV after a few nanoseconds \cite{antici2013modeling, dubois2014target}. These temperature estimates are consistent with the measurements under similar conditions \cite{faenov2015diagnostics, soloviev2017experimental, goriaev2013efficient, faenov1995high} made using a Focusing Spectrometer with Spatial Resolution (FSSR) \cite{faenov2015nonlinear, faenov1997x}.

Figure \ref{oneflow} shows a two-dimensional profile of the linear plasma density obtained using femtosecond optical interferometry 8 ns after irradiation. For reconstructing plasma density details see \cite{soloviev2024research, IDEA_Hipp_2004}. The plasma flow generated by an intense laser and expanding into a vacuum has a collimated structure with an expansion angle no more than a few degrees, which is significantly different from the quasi-spherical expansion of <<nanosecond>> laser plasma \cite{zemskov2024laboratory, Zemskov2024Rayleigh-Taylor}. The investigation of the mechanisms of the collimation of the {<<femtosecond>>} plasma, in contrast to the <<nanosecond>> plasma, will be the subject of our forthcoming paper.

The dashed lines in Fig. \ref{oneflow} (a) mark the slices presented in Fig. \ref{oneflow} (b,c). The one-dimensional profile of the plasma density (Fig. \ref{oneflow} (b)) along the normal to the target (x-axis) is well approximated by the dependence $N_e = A \times e^{-x/w}$, where $w = 0.5$ mm, $A = 7 \times 10^{18}$ cm$^{-2}$. This profile is in excellent agreement with the scaling $N_e = A \cdot e^{-x/(C_s \cdot t)}$ (Ref. \citep{samir1983expansion}), where $t = 8$ ns, $C_S = 60$ km/s is the ion sound speed corresponding to the temperature $T_e \sim 50$ eV. The speed of the plasma flow front is estimated to be 100 km/s.

\begin{figure*}
 \includegraphics[width=16 cm]{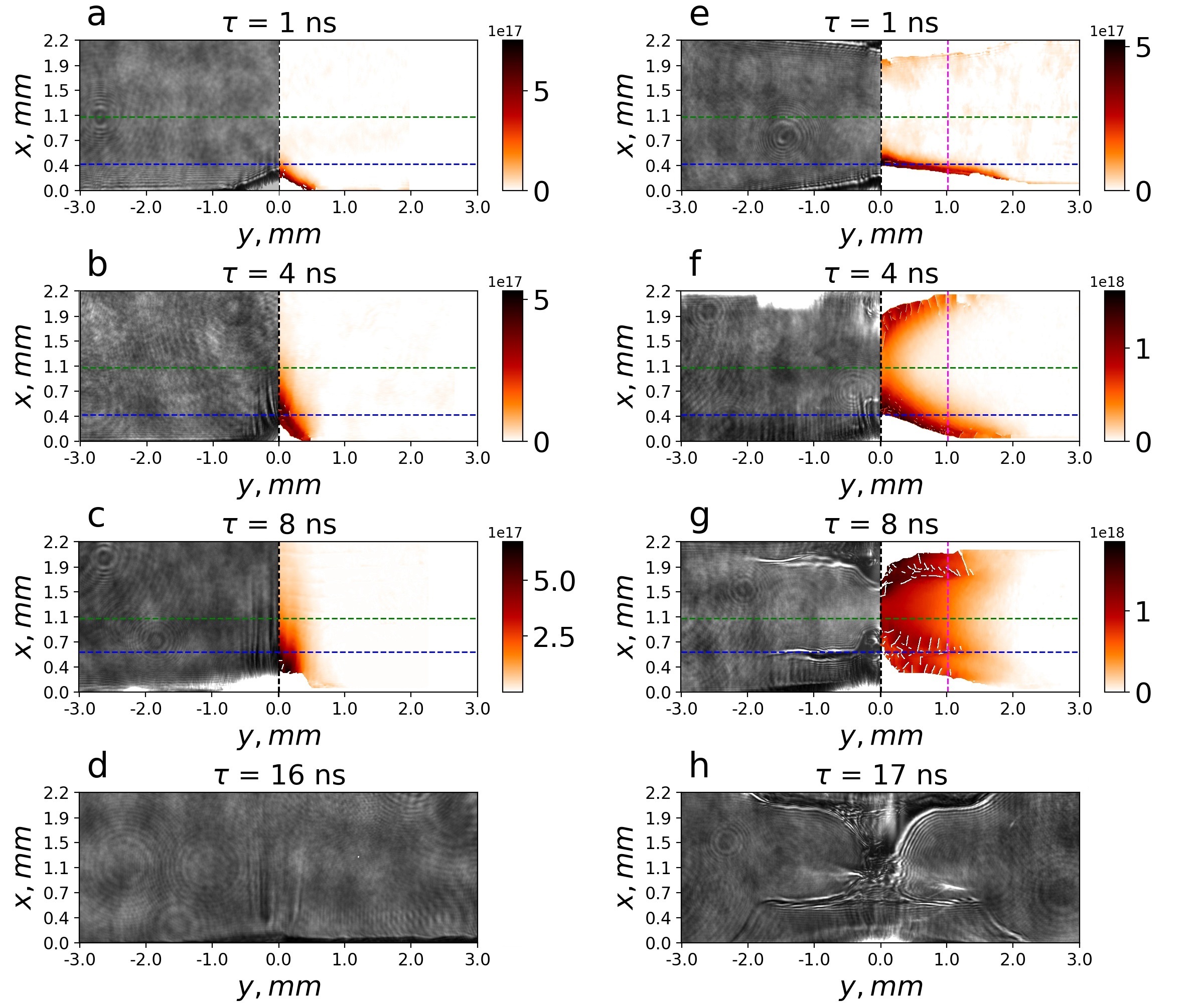}%
 \caption{\label{Compare_NoB} Two-dimensional profiles of linear plasma density. Panels (a, b, c, d) show images of a single flow (no second target and second flow), while panels (e, f, g, h) depict counterpropagating flows of plasma obtained at 1, 4, 8, and 16 (or 17) ns, respectively. }
\end{figure*}  

To perform approximate estimates of the main parameters (Table \ref{Table_in}), it is reasonable to assume that the plasma consists of individual ions H$^+$, C$^+$, and O$^+$. The plasma density at a distance of 1 mm (see Table \ref{Table_out}) of 2.5$\times$10$^{18}$ cm$^{-3}$ was obtained from the linear concentration (Fig. \ref{oneflow}) under the assumption of an axially symmetric flow. According to the FLYCHK database \cite{chung2005flychk}, for the temperature $T_e \sim 50$ eV and a characteristic density of 2.5$\times$10$^{18}$ cm$^{-3}$, the plasma in equilibrium is fully ionized and has an average charge number $\langle Z \rangle = 2$. The average mass number is approximately estimated as $\langle A \rangle = 7$ under the assumption that the plasma composition is determined by the target composition. The initial numerical characteristics of the studied plasma are presented in Table \ref{Table_in} and are used to estimate some parameters characterizing the behavior of the plasma in Table \ref{Table_out}. Estimates for the mean free path of electrons and ions within the flow, as well as ions from different flows, are provided. The mean free path of ions from different flows $l_i(directed)$ = 0.5 mm is calculated using the expression (2) from Ref.\cite{park2012studying} and is of the same order of magnitude as the interaction scale $L$. At the same time, the electron collisions within and between the two flows are significant, as are ion collisions within of the flows. Table \ref{Table_out} presents the values of the Reynolds number $Re = LV/\nu$ and the magnetic Reynolds number $Re_{M} = LV/\eta$, where $L$ is the characteristic spatial scale, $V$ is the flow velocity, $\nu$ is the kinematic viscosity calculated using the expression from Ref.\cite{Ryutov_1999} p. 825, and $\eta$ is the magnetic diffusion coefficient calculated using the formula from Ref.\cite{Ryutov_2000} p. 467.

\begin{table*}
\caption{The values of the initial parameters for each of the two flows.} 
\begin{center}
\begin{tabular}{c|c|c|c|c|c|c|c|c}
\hline 
\hline
\multicolumn{1}{c|}{Parameter} & \multicolumn{1}{c|}{Electron density} & \multicolumn{1}{c|}{Charge state} & \multicolumn{1}{c|}{Mass number}  & \multicolumn{1}{c|}{Flow velocity}  & \multicolumn{1}{c|}{Ion temp.} & \multicolumn{1}{c|}{Electron temp.} & \multicolumn{1}{c|}{Spatial scale} & \multicolumn{1}{c}{Magn. field} 
\tabularnewline

& \multicolumn{1}{c|}{$n_e$, cm$^{-3}$} & \multicolumn{1}{c|}{Z} & \multicolumn{1}{c|}{A}  & \multicolumn{1}{c|}{V, km/s}  & \multicolumn{1}{c|}{T$_{i}$, eV} & \multicolumn{1}{c|}{T$_{e}$, eV} & \multicolumn{1}{c|}{L, cm}   & \multicolumn{1}{c}{B, Т}
\tabularnewline
\hline 

\multicolumn{1}{c|}{Value} & 5 $\times 10^{18}$ & 2 & 7 & 100 & 50 & 50 & 0.1 & 30*
\tabularnewline
\hline

\end{tabular}
\end{center}
\label{Table_in}
\end{table*}

\section{\label{sec:Res}Results}

\begin{figure}
 \includegraphics[width=9 cm]{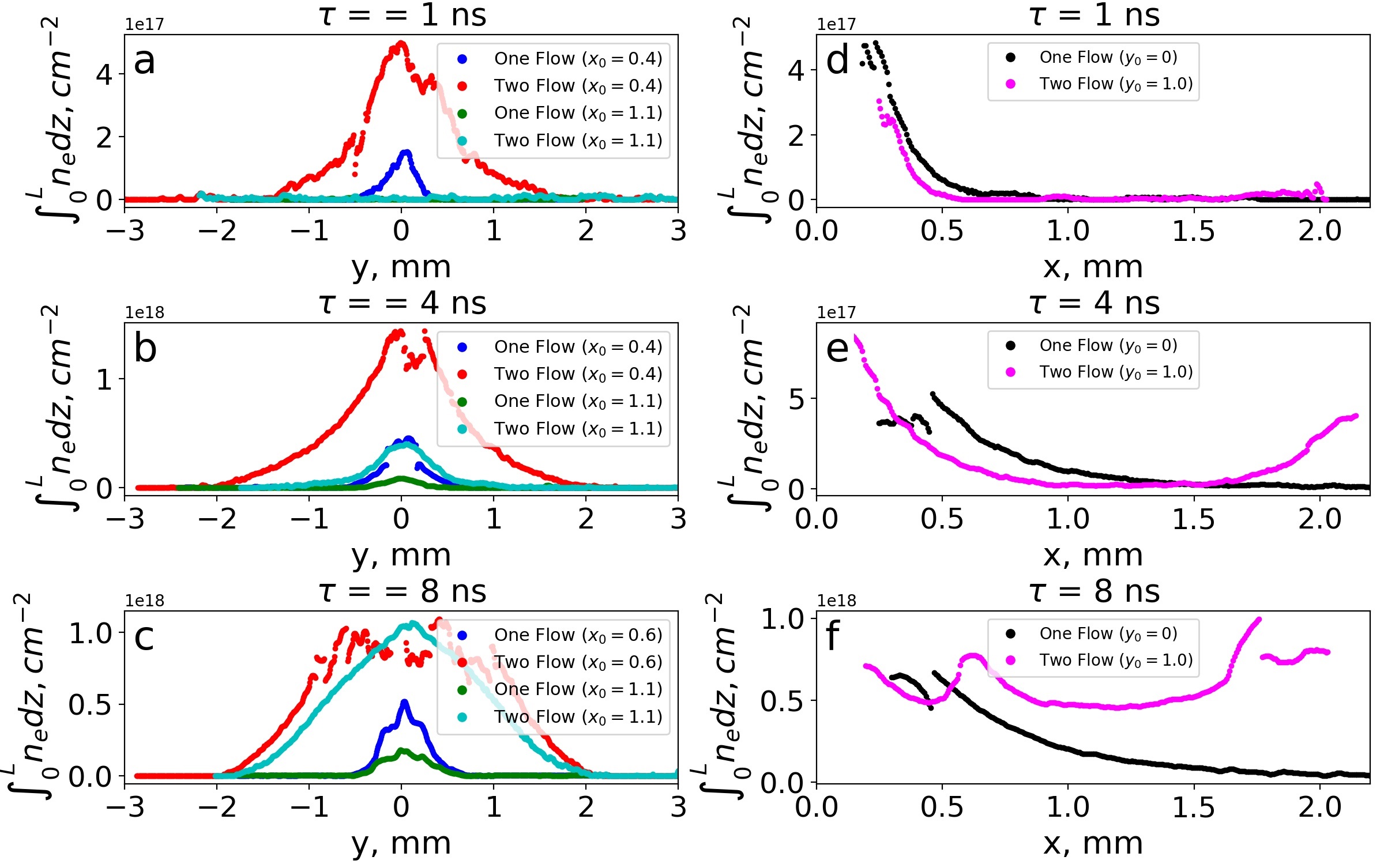}%
 \caption{\label{Slices_NoB} One-dimensional profiles (slices) of linear plasma density. Positions of slices in two-dimensional profiles are marked by dashed lines in Fig. \ref{Compare_NoB}.}

\end{figure}


Shadowgraphy images and two-dimensional plasma density profiles for a single flow (a, b, c, d) and for two counterstreaming flows (e, f, g, h) are presented in Fig. \ref{Compare_NoB}. The plasma snapshots were obtained 1 ns (Fig. \ref{Compare_NoB} (a,e)), 4 ns (Fig. \ref{Compare_NoB} (b,f)), 8 ns (Fig. \ref{Compare_NoB} (c,g)), and 16 - 17 ns (Fig. \ref{Compare_NoB} (d,h)) after target  irradiation. The plasma expansion is axially symmetric, and the plasma density profiles in the (z, x) plane do not differ from those in the (y, x) plane and are therefore not presented. Clearly, the structure of the plasma during the counterstreaming flow interaction differs significantly from the structure of a single flow starting from the very first recorded images ($\tau = 1$ ns) in Fig. \ref{Compare_NoB} (a) and (e). The characteristic transverse scale of the plasma at $\tau = 1$ ns, which was 1 mm for a single flow, increased to 2 mm for the counterpropagating flows (see Fig. \ref{Compare_NoB} (a) and (e) and the slice at $x = 0.4$ mm in Fig. \ref{Slices_NoB} (a)). At the same time, the measured plasma density in the center of the interaction region is below the detectable threshold of our interferometer ($N_{thd} \sim 10^{17}$ cm$^{-3}$). We assume that the reason for the increase of the transverse scale of the plasma flow at this stage ($\tau < 1$ ns)  may be due to additional ionization by energetic electrons and/or X-rays from the opposite targets.




At the next stage ($\tau = 4$ ns), as seen in Fig. \ref{Compare_NoB} (b) and (f), plasma is detected in the  $x = 1.1$ mm plane, i.e. in the middle of the interaction region. Density slices at $x = 1.1$ mm and $x = 0.4$ mm are presented in Fig. \ref{Slices_NoB} (b), while the slice at $y = 0$ is shown in panel (e). One can see that the transverse size (at $x = 0.4$ mm) for the counterpropagating flows increased to 3 mm, while the transverse size of the single flow did not change significantly and remained approximately 1 mm.

At $\tau = 8$ ns (Fig. \ref{Compare_NoB} (c) and (g)), two local density compression layers are detected in the counterstreaming plasma flows, with centers located at approximately $x \approx 0.6$ mm and $x \approx 1.8$ mm. In the region between these compressions and the targets, a less dense plasma is observed. This is demonstrated by the slice for $y = 1$ mm (Fig. \ref{Slices_NoB} (f)), where a double-humped plasma density profile for the two flows is detected, along with an exponentially decaying profile for the single flow. In the density slice for $x = 0.6$ mm, the characteristic transverse scale of two colliding plasma flows reaches 3.8 mm.
This topology indicates stagnation and radial redirection of the plasma flows. 

We conjecture that these plasma compressions are caused by the deceleration of the plasma by the compressed and enhanced toroidal magnetic field (see the simulation in section \ref{sec:Sym}). The existence of a strong  magnetic field is tested by our polarimetric measurements. The details of the polarimetric diagnostic setup and the methods for recovering the depolarization angle are described in the Supplemental materials and in our review \cite{soloviev2024research}. The flaws in the profile of our probing laser pulse and significant scattering at sharp plasma gradients prevented accurate polarimetric reconstruction of the magnetic field structure. However, in several localized areas where the influence of parasitic depolarizing scattering at sharp plasma gradients is minimal, detectable depolarization was measured with a characteristic angle of about $\sim 1^\circ$. With a characteristic linear density of about $10^{18} \, \text{cm}^{-2}$ (Fig. \ref{Compare_NoB}), such depolarization can hint to the presence of magnetic fields in this region with an amplitude on the order of 20 - 30 T.


At the next stage $\tau = 16-17$ ns, as shown in the shadowgraphy image (Fig. \ref{Compare_NoB} (h)), the central cores (<<head>>) of the counterstreaming plasma flows collide in the vicinity of the $y = 0$ axis. Meanwhile at the periphery, where $ r = \sqrt{(y^2 + z^2)} > 0.5 $ mm, the compressed front of the plasma flow has not practically shifted over 9 ns with respect to Fig. \ref{Compare_NoB} (g), indicating the stagnation of the plasma flow in this region.

\begin{table}
\caption{Calculated values of some parameters.} 
\begin{center}
\begin{tabular}{c|c}

\hline
\hline 
\multicolumn{1}{c|}{$\rho$ $[g.cm^{-3}]$} & $5.9\times10^{-5}$  \tabularnewline
\multicolumn{1}{c|}{C$_{S}$ $[km.s^{-1}]$}  & $58$   \tabularnewline
\multicolumn{1}{c|}{V$_{A}$ $[km.s^{-1}]$}  & $110$  \tabularnewline
\multicolumn{1}{c|}{l$_{ee}$ $[cm]$} & $1.25\times10^{-4}$  \tabularnewline
\multicolumn{1}{c|}{$\tau_{col\:ee}$ $[ns]$}  & $1.65\times10^{-3}$  \tabularnewline
\multicolumn{1}{c|}{R$_{Le}$ $[cm]$}  & $5.6\times10^{-5}$  \tabularnewline
\multicolumn{1}{c|}{f$_{ce}$ $[Hz]$} & $8.4\times10^{11}$ \tabularnewline

\multicolumn{1}{c|}{l$_{ei}$ $[cm]$} & $5\times10^{-6}$ \tabularnewline
\multicolumn{1}{c|}{$\tau_{col\:ei}$ $[ns]$}  & $1.9\times10^{-3}$ \tabularnewline

\multicolumn{1}{c|}{l$_{ii}$ $[cm]$} & $8.7\times10^{-5}$  \tabularnewline
\multicolumn{1}{c|}{$\tau_{col\:ii}$ $[ns]$} & $3.3\times 10^{-3}$  \tabularnewline

\multicolumn{1}{c|}{l$_{i}$ (directed) $[cm]$} & $5\times10^{-2}$  \tabularnewline

\multicolumn{1}{c|}{R$_{Li}$  (directed) $[cm]$} & $1.2\times10^{-2}$ \tabularnewline
\multicolumn{1}{c|}{f$_{ci}$ $[Hz]$} & $1.3\times10^{8}$  \tabularnewline
\multicolumn{1}{c|}{M} & $1.7$  \tabularnewline
\multicolumn{1}{c|}{M$_{alf}$} & $0.9$  \tabularnewline
\multicolumn{1}{c|}{$\eta$ $[cm^2/s]$} & $3.4 \times 10^4$  \tabularnewline
\multicolumn{1}{c|}{Re$_{M}$} & $30$  \tabularnewline
\multicolumn{1}{c|}{Re} & $20$  \tabularnewline
\multicolumn{1}{c|}{Eu} & $2.2$  \tabularnewline
\multicolumn{1}{c|}{$\beta$} & $3 \times 10^{-2}$  \tabularnewline
\hline 

\end{tabular}
\end{center}
$^i$ The following symbols are used: $\rho$ — mass density, $C_{S}$ — sound speed, $V_{A}$ — Alfvén speed $(V_{A} = B/\sqrt{4\pi n_{i} m_{i}}$, where $n_{i}$ is the ion density and $m_{i}$ is the ion mass), $l_{ee}$ — mean free path for e-e collisions, $\tau_{col\:ee}$ — time between electron collisions, $R_{Le}$ —  electron Larmor radius, $f_{ce}$ — electron gyrofrequency, $l_{ii}$ — mean free path for i-i ion collisions, $\tau_{col\:ii}$ — ion collision time, $R_{Li}$ — ion Larmor radius, $f_{ci}$ — ion gyrofrequency, $M$ — Mach number, $M_{alf}$ — Alfvén Mach number $(M_{alf} = V/V_{A})$, Re$_{M}$ — magnetic Reynolds number $(\text{Re}_{M} = LV/\eta)$, $\eta$ — magnetic diffusivity, Re — Reynolds number $(\text{Re} = LV/\nu$, where $\nu$ is the kinematic viscosity), Eu = $\sqrt{\frac{\rho V^2}{p}}$ — Euler number, and $\beta = \frac{p}{\frac{B^2}{8\pi}}$ — plasma beta.
\label{Table_out}
\end{table}

\section{\label{sec:Sym} Simulation}

\begin{figure}
 \includegraphics[width=9 cm]{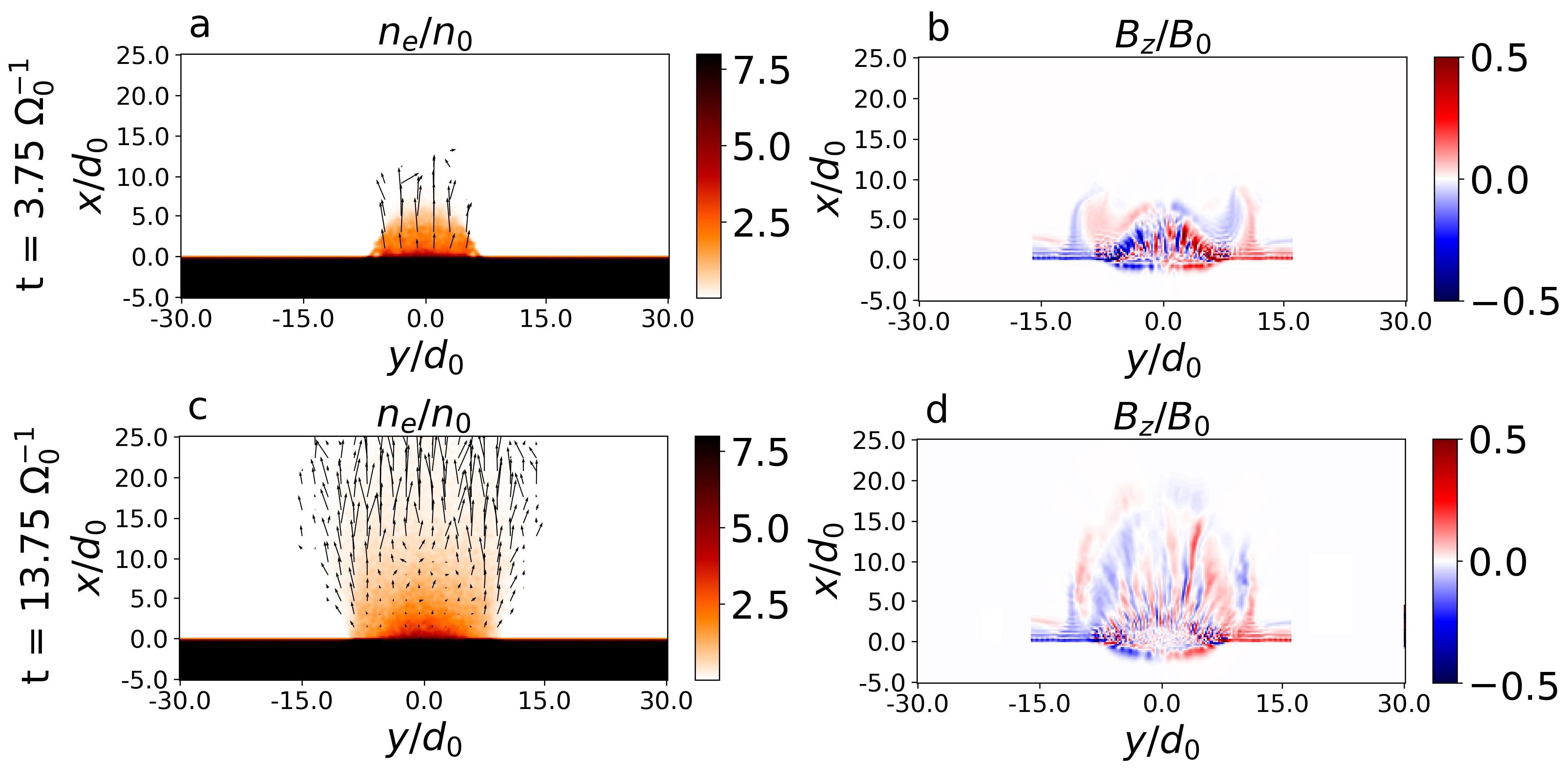}%
\caption{\label{fig:AKA_OneFlow_2d} Results of hybrid code AKA52 2D3V simulations for a single plasma flow. Panels (a) and (c) show two-dimensional profiles of electron density of plasma at times 3.75$t_0$ and 13.75$t_0$, respectively. Arrows indicate directions of ion motion, and arrow length demonstrates relative magnitude of ion velocity. Panels (b) and (d) present two-dimensional profiles of azimuthal magnetic field (for 2D modeling $B_z$) at times 3.75$t_0$ and 13.75$t_0$, respectively. Spatial scales are normalized to ion inertial length $d_0 = c/\omega_{pi} = 19$ $\mu$m, magnetic field is normalized to $B_0 \sim 100$ T, density is set to $n_0 = 2.5 \times 10^{20}$ cm$^{-3}$, and time scales are normalized to inverse ion cyclotron frequency $t_0 = \Omega_{0}^{-1} = \frac{c*M_i}{q_i B_0} \sim 2.3$ ns.}
\end{figure}

\begin{figure}
 \includegraphics[width=9 cm]{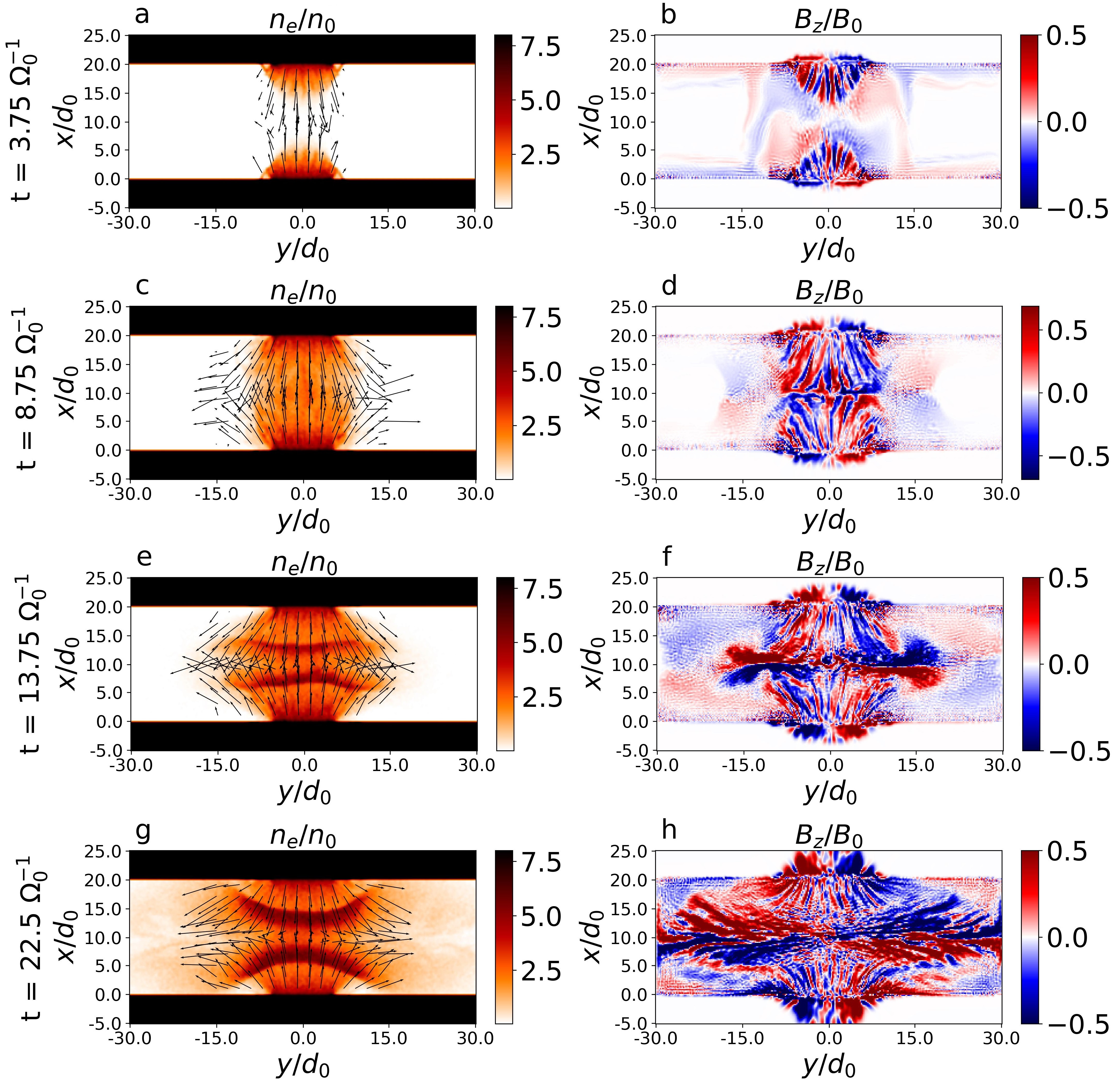}%
\caption{\label{fig:AKA_2d} Results of hybrid code AKA52 2D3V simulations for two counterstreaming plasma flows. Panels (a, c, e, g) show two-dimensional profiles of electron density of plasma at times 3.75$t_0$, 8.75$t_0$, 13.75$t_0$, and 22.5$t_0$, respectively. Arrows indicate directions of ion motion, and arrow length demonstrates relative magnitude of ion velocity. Panels (b, d, f, h) present two-dimensional profiles of azimuthal magnetic field (for 2D modeling $B_z$) at times 3.75$t_0$, 8.75$t_0$, 13.75$t_0$, and 22.5$t_0$, respectively. Spatial scales are normalized to ion inertial length $d_0 = c/\omega_{pi} = 19$ $\mu$m, magnetic field is normalized to $B_0 \sim 100$ T, density is set to $n_0 = 2.5 \times 10^{20}$ cm$^{-3}$, and time scales are normalized to the inverse ion cyclotron frequency  $t_0 = \Omega_{0}^{-1} = \frac{c*M_i}{q_i B_0} \sim 2.3$ ns.}
\end{figure}

To explain the nature of the observed phenomenon, we present numerical simulations of the plasma using a hybrid (PIC - fluid) approach. We did not aim to perform a one-to-one simulation of the experiment due to computational resource limitations. However, our simulations are capable of reproducing important energy relationships (temperatures, kinetic energy) stored in the plasma and the self-generated magnetic field.

We use the hybrid code Arbitrary Kinetic Algorithm (AKA) \cite{akaGitHub} based on the common and well-accepted principles of the previous codes \cite{Winske2003}, such as Heckle \cite{Smets2011}, with advanced features such as an ablation operator and a six-component electron pressure tensor. In the numerical model, ions are described according to the Particle-In-Cell (PIC) formalism, while electrons are described by a ten-moment fluid model: density ($n_e$ equal to the total ion density due to neutrality), mass velocity ($V_{ex}$; $V_{ey}$; $V_{ez}$), and a six-component electron pressure tensor ($\widehat{P_e}_{xx}$; $\widehat{P_e}_{yy}$; $\widehat{P_e}_{zz}$; $\widehat{P_e}_{xy}$; $\widehat{P_e}_{yz}$; $\widehat{P_e}_{zx}$). Electromagnetic fields are considered in the low-frequency approximation (Darwin approximation), neglecting the displacement current. The generalized Ohm law contains three terms: (i) $\mathbf V_i \times \mathbf B$, where $\mathbf V_i$ is the mass velocity of ions; (ii) $(\mathbf J \times \mathbf B)/en$, the Hall effect, where $\mathbf J$ is the total current density that equals the rotor of $\mathbf B$, this term describes the decoupling of ions from the magnetic field; and (iii) $(\boldsymbol{\nabla} \cdot \mathbf P_e)/en$, the divergence of the electron pressure tensor, representing the contribution of the electron fluid \cite{sladkov2021}. Ion collisions are accounted for using the Takizuka-Abe binary collision model \citep{takizuka1977}.

The magnetic field and density in the model are normalized to $B_0 = 100$ T and $n_0 = 2.5 \times 10^{20}$ cm$^{-3}$, respectively, from which follows the normalization of the other quantities. The density determines the ion inertial length, $d_{i0} \sim 19 \, \mu m$, which sets the length scale. The box has a 2D rectangular shape with dimensions $(30 \times 60 \times 1 d_0)$. The magnetic field determines the time normalization, as the time is normalized to the inverse ion cyclotron frequency, $ t_0 = \Omega_0^{-1}$, which is approximately $2.3$ ns. The velocities are normalized to the Alfvén speed $V_0$ calculated from the normalized quantities $B_0$ and $n_0$. In the simulation, the target density was initialized and maintained at a level of 8$n_0$ (i.e., $2 \times 10^{21}$ cm$^{-3}$). The pixel size is chosen so as to resolve the ion inertial length in the interaction region.

Continuous plasma production resulting from the laser-target interaction is simulated using an ablation operator \cite{sladkov2020}, which accounts for the heating of ions and electrons. The operator of particle production maintains constant target density, simulating a reservoir for a target with solid-state density. This approach is adequate, as we observe in the experiment that after irradiation the target serves as a prolonged (several tens of ns) source of plasma. The heating operator linearly increases the electron pressure in a thin layer near the target surface, creating a pressure gradient along the normal to the target surface; this generates an electric field and accelerates ions, giving rise to plasma expansion. The magnitude of the heating operator is fitted to achieve the desired temperature for electrons ($T^{e,i}_{spot} = 10 T_{0}$, where  $T_{0} \approx $ 50 eV for the chosen parameters). The ablation operator is activated when $t\Omega_0 < 20$, after which pressure pumping is turned off, and particles are loaded with a cold temperature; the diameter of the heated area is 15$d_0$.

Figures \ref{fig:AKA_OneFlow_2d} illustrate the evolution of a single plasma flow. We see that the simulation reproduces the experimentally observed flow quite well. As is clear from the experiment, it is sufficiently collimated along the x axis, which is a result of the action of self-generated toroidal magnetic fields. Besides, similarly to the experiment, filamentation of the magnetic field $B_z$ and consequently of the currents (not shown here) is observed in the plasma flow. This filamentation does not have a significant impact on the stagnation phenomenon under discussion. The mechanisms behind the emergence of this filamentation were partially discussed in Ref.\cite{soloviev2024research} and will be discussed in detail in our upcoming paper.

Figures \ref{fig:AKA_2d} illustrate the evolution of the interaction between two counterstreaming plasma flows. Already at the first presented stage $t = 3.75 t_0$, the electron fluid advects the toroidal magnetic field ($B_z$ in 2D), initially generated near the target, into the interaction region (Fig. \ref{fig:AKA_2d} (b)). At $t = 8.75 t_0$, the bulk plasma from both targets reaches the central plane $x = 10 d_0$ (Fig. \ref{fig:AKA_2d} (c)); as a result of the advection, the toroidal magnetic field in this region is compressed and increased (Fig. \ref{fig:AKA_2d} (d)). Further strengthening of the field (see Fig. \ref{fig:AKA_2d} (f) at $t = 13.75 t_0$) leads to the redirection of the plasma in the radial direction, as demonstrated by the arrows indicating the particle motion in Fig. \ref{fig:AKA_2d} (e). Current layers maintaining such a magnetic field structure are formed in the central plane (not shown). In Fig. \ref{fig:AKA_2d} (g), stagnation of the plasma flows and their subsequent redirection in the radial (y in 2D) direction is observed. Since the magnetic field frozen into the plasma is advected with flows, the magnetic layer also expands in the radial direction (Fig. \ref{fig:AKA_2d} (h)). Like in the experiment, regions with density jumps are observed (Fig. \ref{fig:AKA_2d} (g)), and the topology of the flows and magnetic fields matches the topology of the experimentaly observed plasma (Fig. \ref{Compare_NoB} (g, h)).

\section{\label{sec:Discus}Discussion}

The results of numerical simulations of the experiment demonstrate that the self-generated toroidal magnetic field is a crucial component of the interaction of the counterstreaming plasma flows. In hybrid numerical calculations, the magnetic field is generated by the electron fluid according to the Biermann battery mechanism $\sim [\nabla T_e \times \nabla n_e]$ (Refs.\cite{doi2011generation, biermann1950ursprung}). The temperature and density gradients are the largest in the plasma generation region; therefore, the magnetic field is predominantly generated in the vicinity of the target and is advected into the interaction region by two counterstreaming flows. The Reynolds numbers $Re = LV/\nu$ and the magnetic Reynolds number $Re_{M} = LV/\eta$ significantly exceed unity (see Table \ref{Table_out}), indicating that the magnetic field is substantially frozen in the plasma flows.

The advection of the Biermann magnetic field by two counterstreaming plasma flows was phenomenologically described in Ref.\cite{ryutov2013magnetic}. In Ref.\cite{kugland2013visualizing}, this model was used to explain the caustics experimentally observed in proton radiography images. It was claimed that these caustics may arise from the redirection of diagnostic protons by toroidal magnetic fields with induction on the order of 10 T advected to the interaction region by the counterstreaming flows. However, plasma stagnation and its redirection in the radial direction were not observed in those experiments. The authors of Ref.\cite{kugland2013visualizing} believed that this was the result of the weak magnetization of ions of the supersonic (1000 km/s) laser plasma.

In our case, the plasma flow has a lower velocity of about 100 km/s. Simple estimates show that the magnetic field capable of affecting the directed motion of the plasma ($B^2/8 \pi \sim \rho V^2/2$) must have a magnitude on the order of 20-30 T. In the simulations, such a strong magnetic field is predominantly observed in the narrow layer located in the central plane between two targets. However, the experimental data, namely, the wider distance between the regions of compressed density bumps, indicate that the area containing a strong magnetic field may occupy a larger volume.


It is worth noting that in the simulations the total (integrated over the box) energy of the magnetic field throughout the plasma region is significantly lower than the kinetic energy of the plasma flows. The simulations indicate that the Biermann magnetic field generated in the vicinity of the target and subsequently advected and intensified in the interaction region is sufficient to redirect the plasma. Thus, the effect of stagnation and redirection of the plasma flows is based on the redistribution of the energy density of the magnetic field in a thin layer, where the magnetic field is intensified and compressed by the two counterpropagating flows. But in our experiment, where plasma is generated by the action of ultra-high-power laser radiation, the toroidal magnetic fields can be generated not only by the Biermann battery but also by the fountain energetic electrons. Еnergetic electrons receive a significant portion of the laser energy during intense femtosecond laser heating \cite{antici2013modeling, dubois2014target, soloviev2017experimental}. Therefore, fountain fields, as reported in the works \cite{shaikh2016megagauss, borghesi1998large, gopal2008temporally}, can exceed the Biermann fields and reach several hundred tesla. Such high toroidal fields will undoubtedly not disrupt the observed effect; instead, they will lead to a more significant and rapid stagnation of the plasma. We suppose that this is the reason for some discrepancy in the spacing (scales) and timing of plasma stagnation between the experiment and the simulation. It is unfeasible to take the fountain magnetic field into account in numerical modeling with a hybrid code. The influence of the fountain magnetic field on stagnation may be determined by PIC modeling; however, this would require significantly more computational resources and will be addressed in our future studies.

\section{\label{sec:Conclude}Conclusion}

Experiments with counterstreaming flows of laser plasma are of great importance for explaining the nature of many phenomena in astrophysics and plasma physics. We have experimentally observed the stagnation of counterstreaming laser plasma flows induced by short laser pulses with an intensity $ I \sim 2 \times 10^{18} \, W/cm^2$. We have shown by hybrid modeling that the stagnation of the plasma is associated with the advection of a toroidal magnetic field generated by the plasma near the target. The advection of the magnetic field by the plasma into the central region leads to the compression and intensification of the magnetic field in the interaction area of the counterstreaming flows. The simulation demonstrates the enhancement of the magnetic field in the central region by more than 10 times compared to the field observed during the expansion of a single flow, which is in a good agreement with the analytical model presented in the work\cite{ryutov2013magnetic}. The intensified magnetic field results in the redirection of the plasma flow in the radial direction, which is confirmed by the topology of the plasma, observed both in the experiment and the simulation. 

\begin{acknowledgments}
    The work was supported by the 10-th project of the National Center for Physics and Mathematics (NCPhM) “Experimental laboratory astrophysics and geophysics”.

\end{acknowledgments}

\hfill

\textbf{AUTHOR DECLARATIONS}

\textbf{Conflict of Interest}

The authors have no conflicts to disclose.

\hfill

\textbf{DATA AVAILABILITY}

The data that support the findings of this study are available from the corresponding author upon reasonable request.

\bibliography{Stagnbib}

\end{document}